\documentclass[twocolumn,aps]{revtex4}
\usepackage{graphicx}

\begin{document}
\setcounter{page}{58}
\normalsize
\title{Three Dimensional Raman Cooling using Velocity Selective Rapid Adiabatic Passage}

\author{Axel Kuhn, H\'el\`ene Perrin, Wolfgang H\"{a}nsel and Christophe Salomon}

\affiliation{Laboratoire Kastler Brossel, \'{E}cole Normale Sup\'{e}rieure,
24 rue Lhomond, 75231 Paris Cedex 05, France}

\begin{abstract}
We present a new and efficient implementation of Raman cooling of trapped atoms. It uses  Raman pulses with an appropriate frequency chirp to realize a velocity selective excitation through a rapid adiabatic passage. This method allows to address in a single pulse a large number of non zero atomic velocity classes and it produces a nearly unity transfer efficiency. We demonstrate this cooling method  using cesium atoms in a far-detuned crossed dipole trap. Three-dimensional cooling of $1 \times 10^{5}$ atoms down to $2~\mu$K is performed in 100~ms. In this preliminary experiment the  final atomic density is $1.3\times 10^{12}$~at/cm$^3$ (within a factor of 2) and the phase-space density increase over the uncooled sample is 20.  Numerical simulations indicate that temperatures below the single photon recoil temperature should be achievable with this method.

\textbf{Reference}: OSA TOPS on Ultracold Atoms and BEC,
Vol.\,7 (Keith Burnett ed.) 1996, p.58
\end{abstract}
\maketitle

\section*{Key Words}
Trapped atoms, Laser cooling, Ultracold atoms, Coherent optical effects

\section{Introduction}

The recent observation of Bose-Einstein condensation in dilute alkali
vapors was a major advance in the field of atom cooling and trapping
\cite{Cornell,Ketterle}.  The quantum degeneracy thres\-hold was reached by
evaporative cooling of the atomic sample in a magnetic trap. Despite the
tremendous success of these experiments, one can see two important
drawbacks to this method for future applications of these Bose
condenstates: (i) the strong magnetic field gradients required for the
evaporation and their time dependence are
 not easily compatible with the high precision measurements of,
for instance, cold atoms frequency standards and atom interferometers.
(ii) The evaporation is relatively slow (tens of seconds),
 requires a very low rate of inelastic collisions and  leads to a
 substantial loss of atoms (typically a factor $100$). A possible
route to solve both of these problems is to trap the atoms with optical
fields and to apply to the trapped atoms the subrecoil cooling methods
 first developped for free
atoms. Optical fields are easily switched on and off and,
 if properly designed, subrecoil cooling presents
 a priori no loss of atoms. Recently Raman cooling
of trapped Na atoms has produced a sample at $1~\mu$K$=0.42~T_R$, where
$T_R=\hbar^2 k^2/Mk_B$ is the single photon recoil temperature
\cite{Lee96}. This is
 a factor 320 increase in phase-space density but still a factor 300
short of the condensation threshold.

In this paper, we present a new Raman method for subrecoil cooling of
trapped atomic samples and our efforts to reach the
quantum degeneracy on atomic cesium by purely optical methods.
Contrarily to previous demonstrations
of Raman cooling, we use velocity selective Raman pulses in which
 both the amplitude and the frequency are changed in a controlled way.
This produces a velocity selective rapid adiabatic passage which is
very efficient (transfer efficiency close to one) and which can excite
simultaneously a large number of velocity classes.

This paper is organized as follows: we first recall the main elements of
Raman cooling (section~II). We then describe the Rapid Adiabatic Passage
(RAP), its velocity selectivity, and optimization of the frequency chirp to
reach an efficiency close to one (section~III). The effect of the pulse
rate in the cooling sequence and of the atomic motion in a crossed dipole
trap is studied using a numerical simulation of the cooling process for
cesium atoms (section~IV). The experimental setup is described in
section~V. The loading of the crossed dipole trap from a Magneto Optical
Trap (MOT) and the results of Raman cooling using the RAP method are given
in section~VI.

\section{Raman Cooling Scheme}
Raman cooling was first proposed for subrecoil cooling of free atoms in one
dimension \cite{Kasevich92}. It has recently been used to cool cesium atoms down
to a 1D temperature of 3~nK$=T_{\text{R}}/70$ \cite{Reichel95}. Since in
subrecoil cooling schemes, the temperature decreases with increasing
interaction times \cite{Bardou88,Reichel95}, the extension of the method to
trapped atoms is very attractive because of the potentially long storage
time \cite{Lee96}. In addition, in an harmonic trap, the phase space
density scales as $T^{-3}$. In the trap, the cooling can be performed along
a single axis because, under certain circumstances, the motion of the atoms
couples all three degrees of freedom \cite{Lee96}. As we will show in
section \ref{Xsequence}, this coupling causes new effects which lead to a
different cooling strategy than for free atoms.

\begin{figure}
\centerline{\includegraphics[width=60mm]{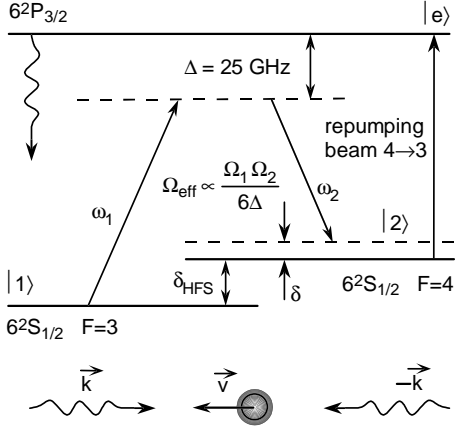}}
\vspace{3mm}
\caption{\footnotesize Raman excitation scheme and subsequent repumping to $F=3$ by
excitation of $F=4\longrightarrow 3$.}
\label{excschem}
\end{figure}

The principle of the cooling mechanism is shown in fig.\ref{excschem} in
the case of cesium atoms. Two counter-propagating Raman pulses excite the
atoms from $F=3$ to $F=4$ in the ground state $6^2S_{1/2}$ and push them by
$2\hbar k$ towards zero velocity. One chooses the pulse shapes and
detunings in a proper way to avoid excitation of atoms at rest and to
select only the desired velocity class.

Subsequently, the excited atoms are brought back to the initial level $F=3$
by pumping them to the excited $6^2P_{3/2}$ state which decays by
spontaneous emission with a rate $\Gamma=2\pi\times 5.3~$MHz. We have
chosen the $F=3$ level in the excited state which decays with a branching
ratio of $2/3$ to the initial state. To avoid a systematic momentum
transfer by this repumping process, the laser shines on the atoms in a
molasses geometry. This repumping process communicates between two and four
photons of random directions. There is no average momentum change in this
process, but this dissipative random walk is required to reach sub-recoil
temperatures. In addition, it produces some heating in the directions
perpendicular to the Raman axis. One has to take this into account to
optimize the cooling sequence (see section~\ref{Xsequence}).

\begin{figure}
\centerline{\includegraphics[width= 70mm]{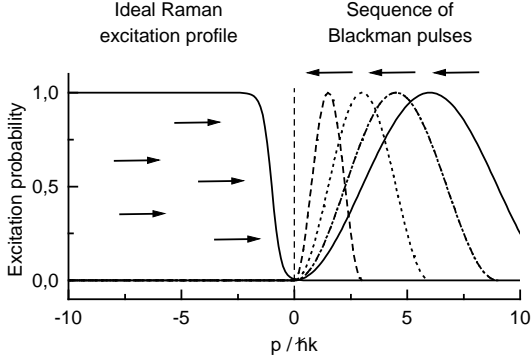}}
\vspace{3mm}
\caption{\footnotesize Idealized Raman excitation profile (shown for $p<0$) and
traditionally realized pulses (for $p>0$)}
\label{sequsch}
\end{figure}

Usually, the initial momentum distribution is much larger than $\hbar k$
and the cooling relies on a repetition of the Raman-- and repumping pulses.
In fig.\ref{sequsch}, we  show an {\em ideal} Raman excitation profile on
the left-hand side.  It provides an excitation probability of `one' for all
atoms at $p<-2 \hbar k$ and tends towards zero around $p=0$. Traditionally,
such an ideal excitation profile was replaced by a sequence of $\Pi$-pulses
at different detunings \cite{Lee96,Kasevich92,Reichel94,Reichel95},
starting at large Raman detuning $\delta$ and compressing the velocity
distribution by a reduction of $\delta$ from pulse to pulse.

 On free atoms,
the direction of the Raman pulses is changed from pulse to pulse to cool
both wings of the initial velocity distribution.In a three dimensional
trap, there is no need to alternate the Raman beam directions because the
atoms oscillate in the harmonic potential. However, the dissipative
repumping process and the oscillatory motion may bring some atoms to high
velocities that are no more affected by the Raman pulse with the smallest
detuning. Therefore, the sequence has to be restarted at higher detuning to
re-collect these atoms. In fact, one single Raman pulse affects only a
small fraction of the velocity distribution (especially the narrow band
pulses at small $\delta$). There are also some technical disadvantages of
such a sequence. One has to fulfill the $\Pi$-pulse condition for each of
them, and the pulse duration has to be adapted to $\delta$ to avoid the
excitation of atoms at rest.

Most of the disadvantages of such a $\Pi$-pulse sequence vanish if one
provides pulses with an ideal excitation profile in momentum space.
In the following section, we  demonstrate the usefulness of the well known
rapid adiabatic passage (RAP) to tailor the desired Raman excitation profile.

\section{Rapid Adiabatic Passage}
\label{RAP}
The first successful combination of stimulated Raman scattering involving
adiabatic passage in a $\Lambda$-type three level system by Bergmann
\cite{stirap} was based on time delayed Raman pulses with fixed
frequencies. The excitation probability of such a scheme reflects the
Fourier transform of the applied pulses. The same holds for the previous
Raman cooling schemes using coinciding Blackman  \cite{Kasevich92} or
square pulses \cite{Reichel95} with fixed frequencies at large detuning
$\Delta$.

In the latter case, the simplification of the three level system to an
effective two level system is justified if $\Delta\gg|\Omega_{1,2}|$.
In our method, we further allow the Raman detuning
$\delta_{\text{eff}}$ to be time dependent. The interaction Hamiltonian is then
\begin{equation}
H_{\text{eff}}(t)=-\frac{\hbar}{2}
\left(\begin{array}{cc}0&\Omega_{\text{eff}}(t)\\ \Omega_{\text{eff}}^*(t)&2 \delta_{\text{eff}}(t)\end{array}
\right).
\label{H2}
\end{equation}
It describes the evolution of $|\Psi\rangle$ in the basis of the
initial level $|1\rangle$  and final level $|2\rangle$, the intermediate level
being eliminated. In this effective two level system,
it is clear that rapid adiabatic passage (RAP), as described by
Loy \cite{Loy}, can be employed when the Raman detuning $\delta_{\text{eff}}$
is chirped across the resonance.

If one takes into account the different magnetic sublevels $m_F$, the
effective Rabi frequency and detuning read
\begin{equation}
\Omega_{\text{eff}}=\frac{\Omega_1 \Omega_2}{6\Delta} \sqrt{1-\left(\frac{m_F}{4}\right)^2}
\end{equation}
\begin{equation}
\delta_{\text{eff}}=\delta  +
\frac{\delta^2_{\text{HFS}}(\Omega_1^2-\Omega_2^2)+
\Delta\,\delta_{\text{HFS}}(\Omega_1^2+\Omega_2^2)}
{4 \Delta (\Delta^2-\delta^2_{\text{HFS}})}
\label{lightshift}
\end{equation}
At this point, we note that it is impossible to have the same $\Pi$-pulse
condition for all magnetic sublevels, because the effective Rabi
frequencies depend on $m_F$. In addition, because of the differential light
shift, the effective detuning $\delta_{\text{eff}}$ depends on $\Omega$.

\begin{figure}
\centerline{\includegraphics[width= 70mm]{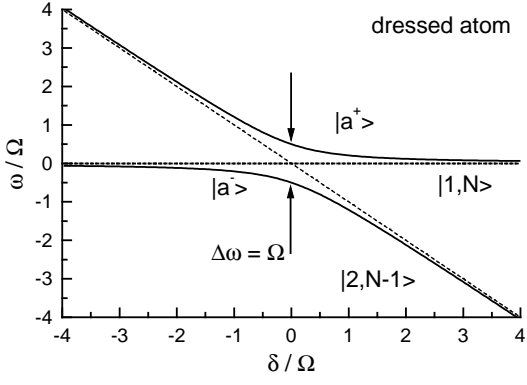}}
\vspace{3mm}
\caption{\footnotesize Avoided crossing at $\delta=0$}
\label{avoid}
\end{figure}

In a dressed state picture, the eigenstates of the two state Hamiltonian
read
\begin{equation}\begin{array}{r}
|a^+\rangle=\cos\phi|1\rangle-\sin\phi|2\rangle\\
|a^-\rangle=\sin\phi|1\rangle+\cos\phi|2\rangle
\end{array}
\qquad
\tan\phi=\frac{\sqrt{\Omega^2+\delta^2}-\delta}{\Omega},
\end{equation}
with the corresponding eigenfrequencies
\begin{equation}
\omega^\pm = -\frac{\delta}{2}\pm\frac{1}{2}\sqrt{\Omega^2+\delta^2}
\end{equation}
Both the mixing angle $\phi$ and the eigenfrequencies $\omega^\pm$ are a
function of the Raman detuning $\delta$. If $\delta$ sweeps across the
resonance, $\phi$ turns from $\pi/2$ to $0$ (which is valid for
$\Omega\ll\delta$ in the wings of a Raman pulse), and by consequence
$|a^-\rangle$ evolves from $|1\rangle$ to $|2\rangle$. In fig.\ref{avoid},
the corresponding avoided crossing at $\delta=0$ is shown.

If the extend of the frequency chirp is limited, only the atoms which
experience such a crossing due to their Doppler shifted resonance frequency
are excited by the Raman pulse. Even atoms which do not experience a
crossing follow $|a^-\rangle$ adiabatically, but the dressed state returns
to the initial state $|1\rangle$ in the end of the pulse.

\begin{figure}
\centerline{\includegraphics[width= 75mm]{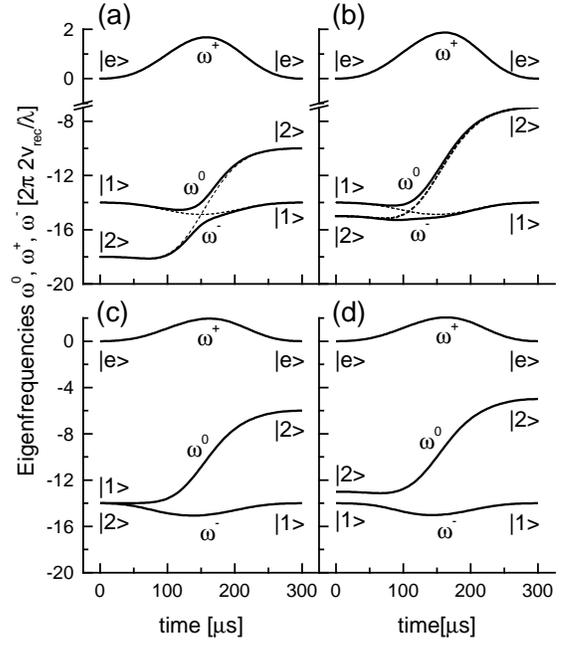}}
\vspace{3mm}
\caption{\footnotesize Energies of the dressed levels (solid lines) in response to a
frequency chirped Blackman pulse with a duration of $300~\mu$s and a pulse
area $\int\Omega(t)dt=3\pi$. The dashed lines represent the energies of the
uncoupled states. The chirp extends from
$\delta=-24\,kv_{\protect\text{rec}}$ to
$\delta=-6\,kv_{\protect\text{rec}}$. Due to the Doppler shift
$\delta_{\protect\text{eff}}=\delta_{\protect\text{chirp}}+2 kv$ of the
counter-propagating Raman beams, atoms in the velocity interval $-12
\protect\lesssim v/v_{\protect\text{rec}} \protect\lesssim -3$ are
affected. The evolution is shown for four different initial velocities --
{\bf (a):} $v/v_{\protect\text{rec}}= -7.5$; {\bf (b):}
$v/v_{\protect\text{rec}}= -4$; {\bf (c):} $v/v_{\protect\text{rec}}= -3$;
{\bf (d):} $v/v_{\protect\text{rec}}= -2$.}
\label{multicross}
\end{figure}

In fig.\ref{multicross}, the evolution of the eigenfrequencies of a real
three level system is shown for a fixed frequency chirp and different
Doppler shifts. (a) shows a crossing right in the center of the chirp, (b)
shows that there is still an avoided crossing when the Doppler shift is
close to the border of (but within) the chirp. If the Doppler shift
coincides with the detuning in the end (c), $|a^+\rangle$ and $|a^-\rangle$
become degenerate and both states are equally populated. (d) shows that
there is no more avoided crossing if the Doppler shifted resonance is not
within the chirp, i.e. there is no Raman excitation in such a case and no
momentum exchange takes place. In fact, the extend of the chirp determines
the range of velocities which are excited.

\subsection{Shape of the $\delta$ chirp}
To insure that the state vector $|\Psi\rangle$ follows the dressed state
$|a^-\rangle$ adiabatically throughout the interaction, the adiabaticity
criterion (see \cite{Messiah})
\begin{equation}
|\langle a^+|\frac{d}{dt}|a^-\rangle| \ll |\omega^+-\omega^-| = \sqrt{\Omega^2+\delta^2},
\end{equation}
needs to be fulfilled. It is most stringent in the vicinity of the crossing,
where it can be expressed as
\begin{equation}
\left|\frac{d}{dt}\delta\right| \ll \Omega^2.
\end{equation}

\begin{figure}
\centerline{\includegraphics[width= 70mm]{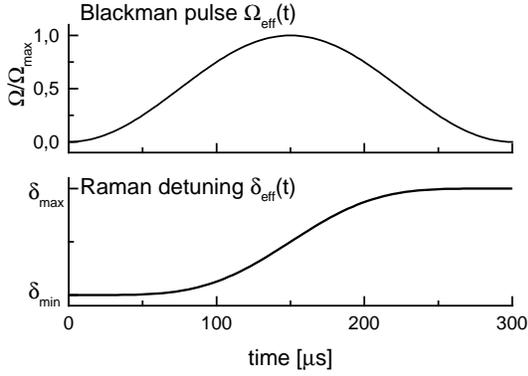}}
\vspace{3mm}
\caption{\footnotesize Form of the Blackman shaped Raman pulse (upper part) and the
superimposed frequency chirp (lower part).}
\label{sweep}
\end{figure}

Due to the different atomic velocities, the exact position of the crossing
within the chirp is not known. Therefore the chirp has to be shaped like
\begin{equation}
\delta(t)\propto\int\Omega^2(t) dt,
\end{equation}
to insure the same adiabaticity condition for all atoms. The chirp thus
reflects the pulse shape and is fast at high Rabi frequencies and slow at
small $\Omega$. Fig.\ref{sweep} shows the shape of the Raman pulse together
with the optimized chirp.

\subsection{Efficiency and Selectivity}
\begin{figure}
\centerline{\includegraphics[width= 70mm]{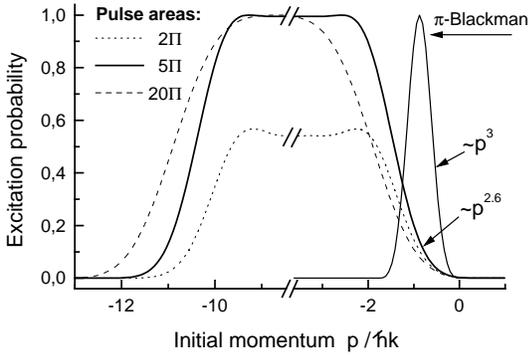}}
\vspace{3mm}
\caption{\footnotesize Transfer efficiency of frequency chirped Raman pulses in comparison
with the excitation profile of a non-chirped Blackman pulse. All pulses
have a duration of $300~\mu$s and are chirped by
$\Delta\delta=2\pi\times74$~kHz, which extends over 9
$v_{\protect\text{rec}}$. The start frequency is shifted from resonance in
order to avoid  exciting atoms at rest. Also shown is the power law
dependence of the excitation profiles near $p=0$.}
\label{transfer}
\end{figure}

A straightforward numerical simulation of the RAP excitation was done
solving the time-dependent Schr\"{o}dinger equation. We show the resulting
transfer efficiency as a function of the atomic velocities (i.e. the
Doppler shift) for different pulse amplitudes in fig.\ref{transfer}.
The simulation shows that a box-like excitation profile is realized
with such pulses, but it reveals at the same time that the border of the
profile is not as steep as the one achieved with a Blackman pulse of
same duration.

Once the recoil limit $p\lesssim\hbar k$ is reached, a description of the
cooling process using L\'{e}vy flight statistics \cite{Levy} shows that
there is an ideal exponent $\alpha$ for the excitation probability
$P_{\text{exc}} \propto v^\alpha$ around $v=0$. It depends on the dimension
$D$ of the cooled atomic ensemble and
\begin{equation}
\frac{D}{\alpha} = \mu \lesssim 1
\end{equation}
is the optimal choice for free atoms \cite{Reichel95}. There is no real
`zero' in the excitation profile of the chirped Raman pulses and one has to
choose an arbitrary cut-off to determine the origin. If one chooses
$P_{\text{exc}}\, t_{\text{trap}}\, \Gamma_{\text{cool}} = 0.25$ as
condition to determine this origin (i.e. a probability of 25\% to excite
atoms at rest during the trapping time $t_{\text{trap}}$). We obtain for a
typical cooling time $t_{\text{trap}}=200$~ms, a cooling pulse rate
$\Gamma_{\text{cool}}=0.5$~ms and pulse parameters  of fig.\ref{transfer}:
\begin{equation}
\alpha_{\text{RAP}} \simeq 2.6 .
\end{equation}
In this case, $D/\alpha$ is slightly greater than one, which is not the
optimum choice to enter far in the subrecoil region. By contrast, a
Blackman pulse without frequency chirp has a first minimum in its
excitation profile where $\alpha_{BM}
\simeq 3.3$ is well determined. Therefore, a combination of both profiles, a chirped Blackman
at large detuning and a narrow fixed frequency Blackman pulse might turn
out to be the optimum choice.

\section{Cooling Sequence}
\label{Xsequence}
In three dimensional Raman Cooling, one has to be aware that the
re-thermalization after each pulse has to take place. We call
$\Gamma_{\text{therm}}$ this thermalization rate and
$\Gamma_{\text{couple}}$ the rate of coupling of the atomic motion to the
Raman axis. If the pulse rate $\Gamma_{\text{cool}}$ in the cooling
sequence is larger than both of these rates, i.e. if
\begin{equation}
\Gamma_{\text{cool}}>\mbox{max}(\Gamma_{\text{therm}},\Gamma_{\text{couple}}),
\end{equation}
the coupling of the atomic motion perpendicular to the Raman axis to this
same axis becomes ineffective.
In such a case, the ensemble is cooled in one dimension at the
expense of heating the other degrees of freedom due to the
dissipative repumping process.

\begin{figure}
\centerline{\includegraphics[width= 60mm]{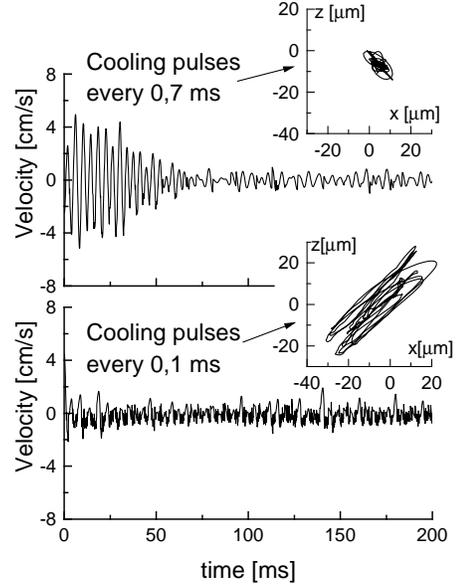}}
\vspace{3mm}
\caption{\footnotesize Time evolution of the velocity component along the Raman axis for different pulse
rates $\Gamma_{\protect\text{cool}}$. Projections of the corresponding
trajectories (between 100~ms and 200~ms)  in the plane containing the
vertical and the Raman axis are shown in the insets.}
\label{coolsim}
\end{figure}

The Monte Carlo simulation of single atom trajectories in fig.\ref{coolsim}
demonstrates this effect. We have calculated 3D classical trajectories with
random initial positions and velocities and have chosen a simplified
excitation profile for the Raman pulses:
\begin{equation}
P_{\text{exc}}(p) = \left\{
\begin{array}{lcl}
1 & & p<-3\hbar k\\
|p/3\hbar k|^2 & \mbox{for} & -3\hbar k\le p \le 0\\
0 & & p>0
\end{array}
\right.
\end{equation}
In the simulation, every $\Gamma^{-1}_{\text{cool}}$ the atom can be
excited according to $P_{\text{exc}}$. Once a Raman excitation takes place,
the repumping process is simulated by adding three momenta $\hbar k$ of
random directions.

For the different cooling rates shown in fig.\ref{coolsim}, the velocity
component along the Raman axis decreases in all cases. However, a closer
look to the trajectory for the high cooling rate
$\Gamma^{-1}_{\text{cool}}=0.1$~ms reveals that the other degrees of
freedom are not cooled efficiently.

\begin{figure}
\centerline{\includegraphics[width= 60mm]{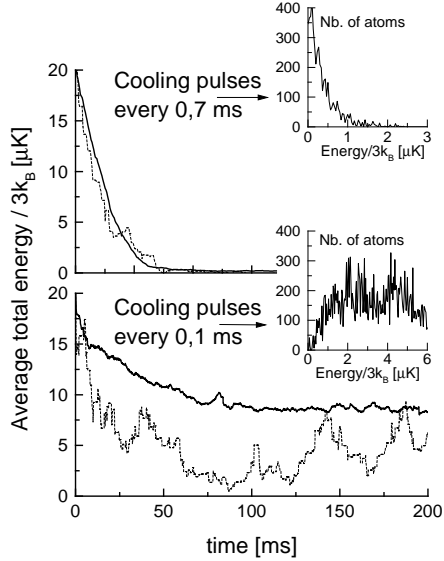}}
\vspace{3mm}
\caption{\footnotesize Total energy loss for different pulse rates $\Gamma_{\protect\text{cool}}$.
The solid lines show the average over 100 atoms, the dashed lines
correspond to  single trajectories. In the insets are shown the energy
distributions in the steady-state.}
\label{coolenergy}
\end{figure}

\begin{figure}
\centerline{\includegraphics[width= 65mm]{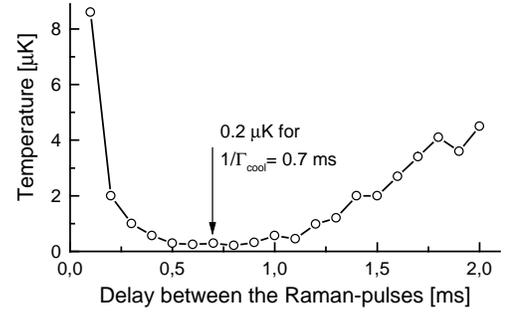}}
\vspace{3mm}
\caption{\footnotesize Temperature reached after 100~ms Raman cooling as a function of
the pulse rate.}
\label{reachedtemp}
\end{figure}

The energy loss due to the different cooling rates in fig.\ref{coolenergy}
underline this effect. After 100~ms, an equilibrium is reached in case of a
high pulse rate, i.e. the heating of the perpendicular motion is balanced
by the combination of cooling and coupling to the Raman axis. The
`temperature' of $9~\mu$K is rather high in this case (there is no thermal
distribution -- we express the average energy in terms of a `temperature').
On the other hand, the low cooling rate $\Gamma^{-1}_{\text{cool}}=0.7$~ms
allows an effective coupling of the atomic motion to the Raman axis.
A simulation of the temperature obtained after a typical cooling time
$t_{\text{trap}}=200$~ms is shown in fig.\ref{reachedtemp} as function of
$\Gamma^{-1}_{\text{cool}}$. Due to the limited cooling time, there is an
optimum choice $\Gamma^{-1}_{\text{cool}}=0.7$~ms for which the temperature
reaches 0.2~$\mu$K, i.e. the recoil temperature for cesium. At lower pulse
rates, the number of pulses is too small to bring the temperature further
down. We have observed that for cooling times longer than 200~ms, the
temperature still decreases below 200~nK.

\section{Experimental Setup}

\subsection{YAG and Raman beams}

\begin{figure}
\centerline{\includegraphics[width=60mm]{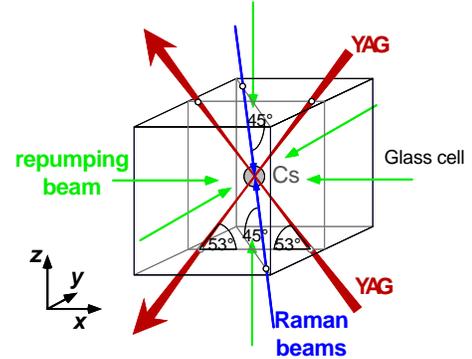}}
\vspace{3mm}
\caption{\footnotesize Configuration of the dipole and Raman beams.}
\label{beamconf}
\end{figure}

Atoms are loaded in the crossed dipole trap from a vapor cell magneto
optical trap (MOT) \cite{Monroe90}. The beam configuration of the red
detuned dipole trap is shown in fig.\ref{beamconf}. We use a 1064~nm
TEM$_{00}$ Nd:YAG laser which is split into two arms having each a power of
7~W. These two beams cross in their focal points with a common waist
$w_0=80~\mu$m. Both of them propagate in a vertical plane and make an angle
of $\pm 37^\circ$ with the vertical.

\begin{figure}
\centerline{\includegraphics[width=65mm]{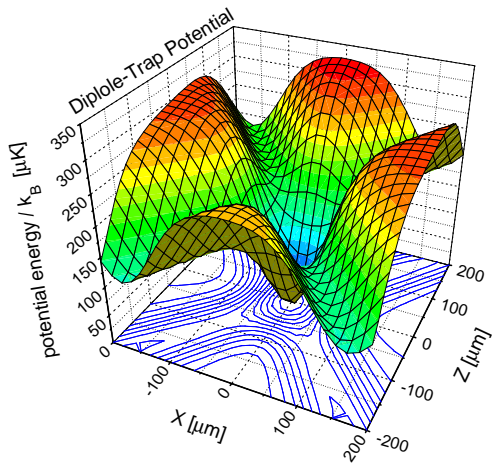}}
\vspace{3mm}
\caption{\footnotesize Cut of the dipole trap potential in the plane of the beams.}
\label{dipolpot}
\end{figure}

The attractive potential (see fig.\ref{dipolpot}) is caused by the ground
state light shift
\begin{equation}
\Delta E = \frac{\hbar\Omega_{\mbox{\tiny YAG}}^2(\vec{r})}{4 \Delta_{\mbox{\tiny YAG}}}
\end{equation}
and corresponds to a well depth of $150~\mu$K.  With these laser
parameters, gravity is compensated only in the intersection volume of the
two beams. Because of the large detuning of the YAG laser, the maximum
photon scattering rate at the center of the trap is 3~s$^{-1}$.  The
calculated oscillation frequencies near the bottom of the potential are
\begin{equation}
\nu_x=458~\mbox{Hz}\qquad\nu_y=577~\mbox{Hz}\qquad\nu_z=350~\mbox{Hz},
\end{equation}
where $x$ and $y$ design the directions in the horizontal plane,
$y$ being perpendicular to both of the dipole beams, and $z$ the
vertical axis.

The Raman cooling beams are spatially filtered using optical fibers and
propagate along the (1,1,1) direction referred to the eigen-axes of
oscillation in the dipole trap. This provides an effective coupling of the
one dimensional cooling to all degrees of freedom of the atomic motion. The
waist  of the Raman beams is on the order of $\sim 1$~mm and their power is
up to 35~mW. For a detuning $\Delta = 2\pi \times 25$~GHz, the maximum
effective Rabi frequency is $\Omega_{\mbox{\tiny eff}} = 4 \times
10^5$~rad.s$^{-1}$. The two beams are issued from two diode lasers whose
frequency difference is phase-locked to a stable frequency source at the
9.2~GHz Cesium hyperfine frequency \cite{Reichel94,Reichel95}.  The shape,
central frequency and frequency chirp of the pulses are achieved using
acousto-optic modulators. The Raman tranfer from $F = 3$ to $F = 4$ is
followed by a repumping pulse, resonant with the
$F=4~\longrightarrow~F$'$=3$ transition which returns the atoms to $F=3$.
This repumping laser illuminates the atoms in a six beam molasses geometry
(fig.\ref{beamconf}).

\subsection{Detection}

We measure the atom number, the velocity distribution of the trapped atoms
and the size of the cloud: the number of atoms and the velocity
distribution are deduced from fluorescence measurements while the size is
obtained by absorption imaging \cite{Ketterle}. After switching off the YAG
beams, we turn on a $\sigma^+$ polarized probe beam tuned to the
$F=4~\longrightarrow~F$'$=5$ transition which overlaps the dipole trap. It
is slightly red detuned and in a standing wave configuration. $2\%$ of the
fluorescence light is collected with a lens on a calibrated photodiode. For
measuring the momentum distribution, we use velocity selective Raman
transition \cite{Kasevich91}. We transfer a narrow velocity class from
$F=3$ to $F=4$ with a $200~\mu$s Blackman $\Pi$--pulse. The fluorescence of
these selected atoms is then recorded as above. By scanning the Raman
detuning $\delta$ we obtain the full momentum distribution, with a
resolution of about $\hbar k/2$.

The absorption image is obtained on a CCD camera by shining a 15~$\mu$s
probe beam pulse on the trapped atoms in the (1,1,0) direction (in the
horizontal plane). The probe is typically detuned by $-1.5\Gamma$ from
resonance. The trap is imaged with a magnification of 4 on the CCD. The
image appears as the ratio between a picture with atoms and a reference
picture without atoms. The absorption picture gives the size of the trap
with a resolution of about $8
\mu$m and an independant measure of the number of atoms.

\section{Experimental Results}

\subsection{Loading the Dipole Trap}

\begin{figure}
\unitlength1mm
\begin{picture}(80,42)
\put(0,2){ \includegraphics[width=38mm]{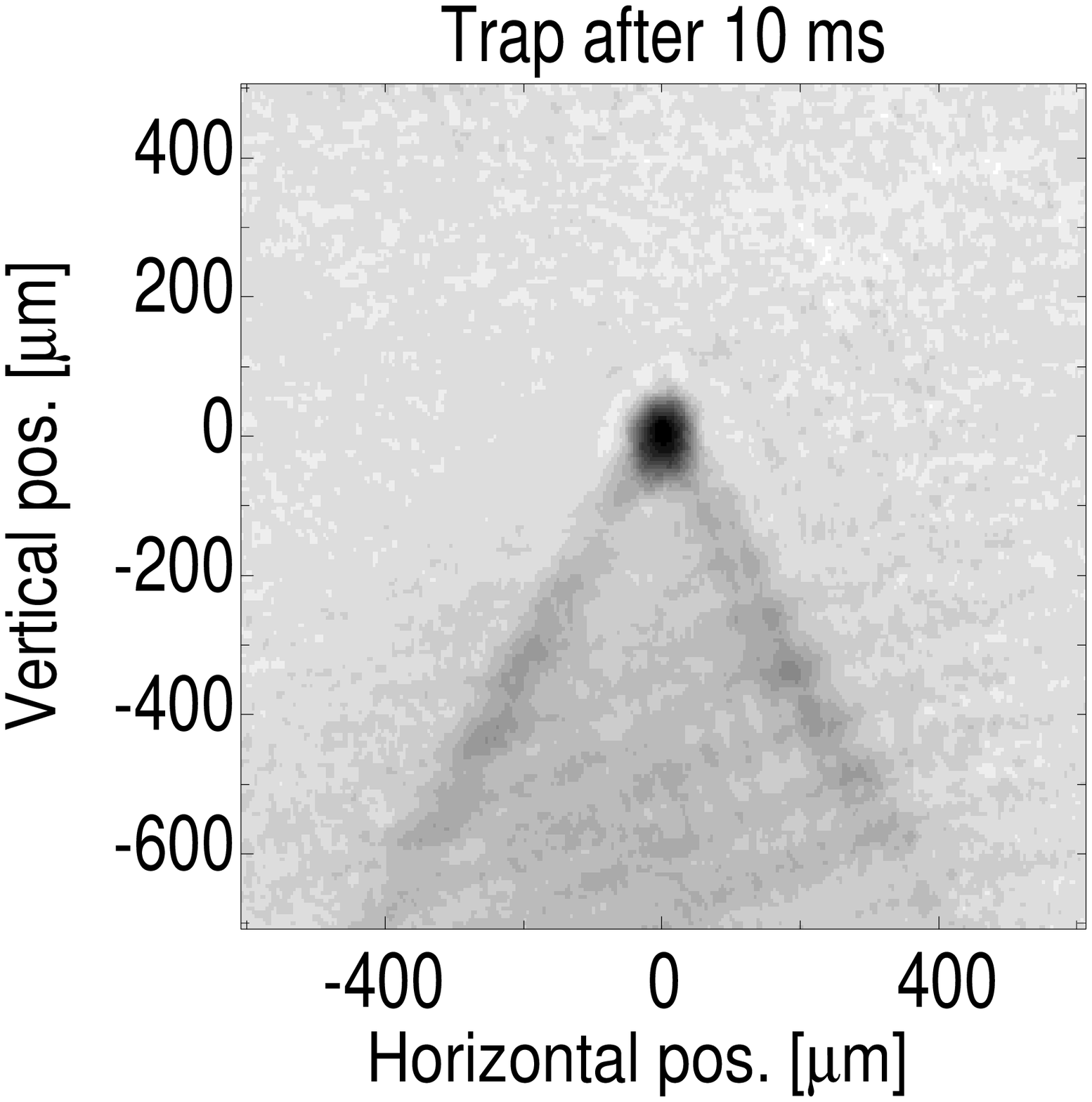}}
\put(43,2){\includegraphics[width=38mm]{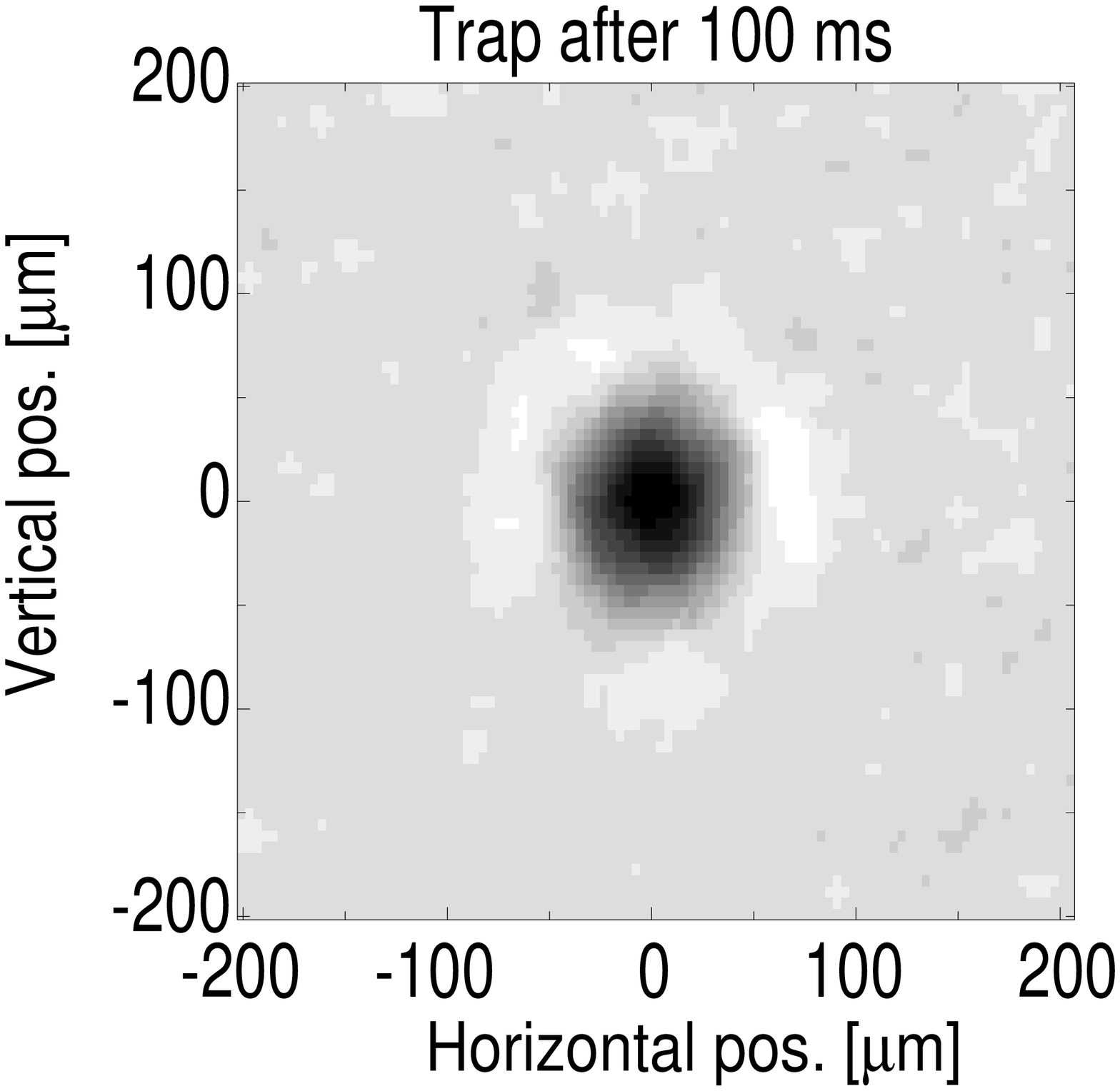}}
\end{picture}
\caption{\footnotesize Absorption images of cesium atoms in the crossed
trap in the lin$\perp$lin configuration of the YAG beams.
After 100~ms, $10^5$ atoms remain in the `harmonic' trap and form a cloud of
Gaussian shape with $\sigma_{x,y}\simeq 24~\mu$m, $\sigma_z\simeq 30~\mu$m.}
\label{linperplin}
\end{figure}

Cesium atoms are first collected for 1~s in the MOT from the background gas
at $10^{-9}$~mbar. In order to capture the largest possible number of
atoms, the YAG trap and the MOT overlap in time for about 100~ms. Within
this time, the MOT is contracted by doubling the magnetic field gradient to
30~Gau\ss/cm and reducing the intensity of the MOT beams by a factor of 20
to 1~mW/cm$^2$. The detuning of the beams is increased to $-10~\Gamma$ for
a short period ($\sim 5$~ms) to further cool  the atoms before switching
the MOT off. Absorption images recorded 10~ms and 100~ms after switching
off the MOT are shown in fig.\ref{linperplin}. After 10~ms the atoms
initially at the intersection of the YAG beams remain trapped while the
others fall preferentially along the YAG beams.  After 100~ms, $10^5$~atoms
($1\%$ of the number of atoms in the MOT) remain in the crossed dipole trap
with a density of $3\times 10^{11}~\mbox{atoms}/\mbox{cm}^3$ (within a
factor 2) and an initial temperature of $6~\mu$K.

\begin{figure}
\unitlength1mm
\begin{picture}(80,42)
\put(0,2){\includegraphics[width=38mm]{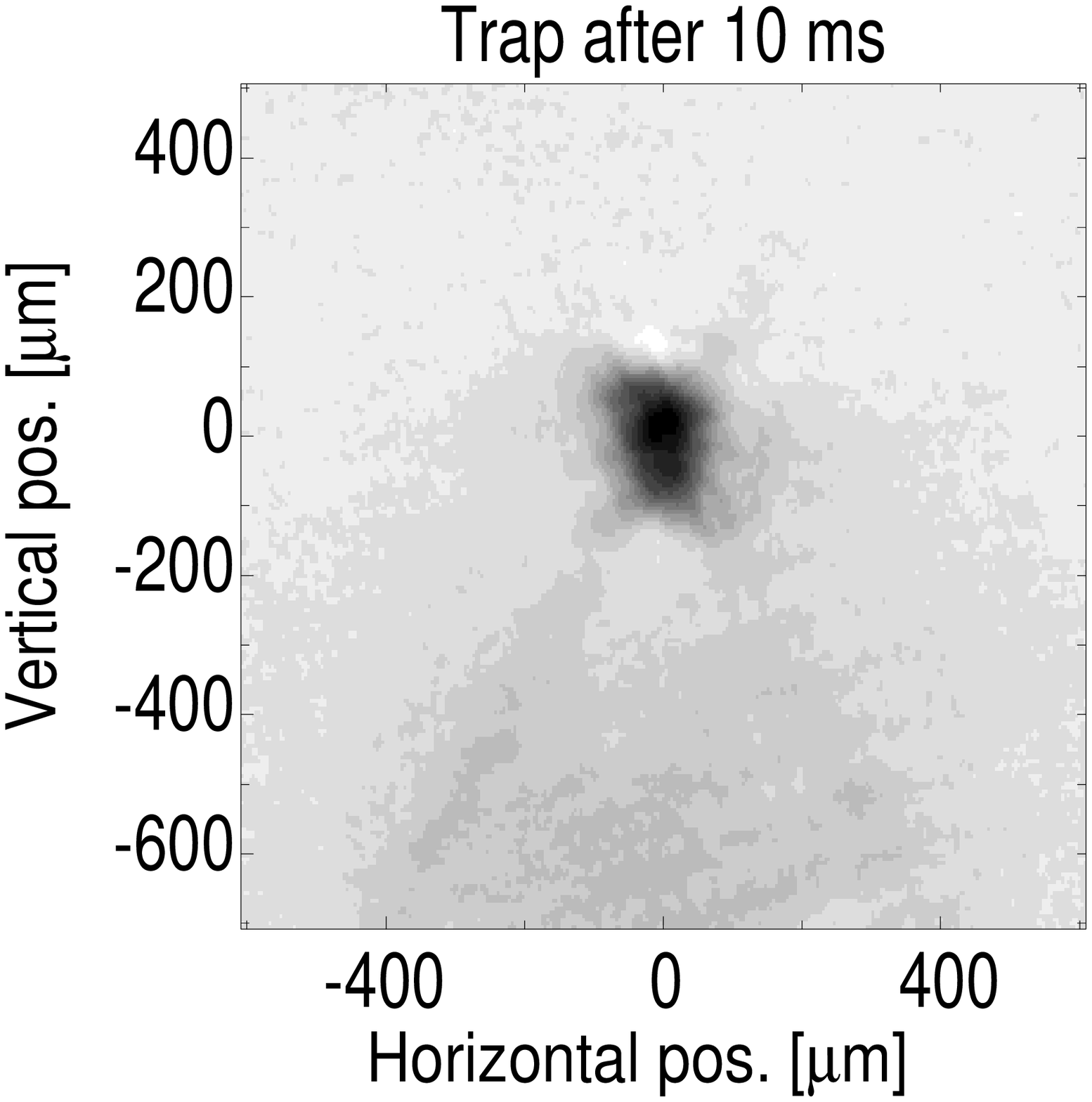}}
\put(43,2){\includegraphics[width=38mm]{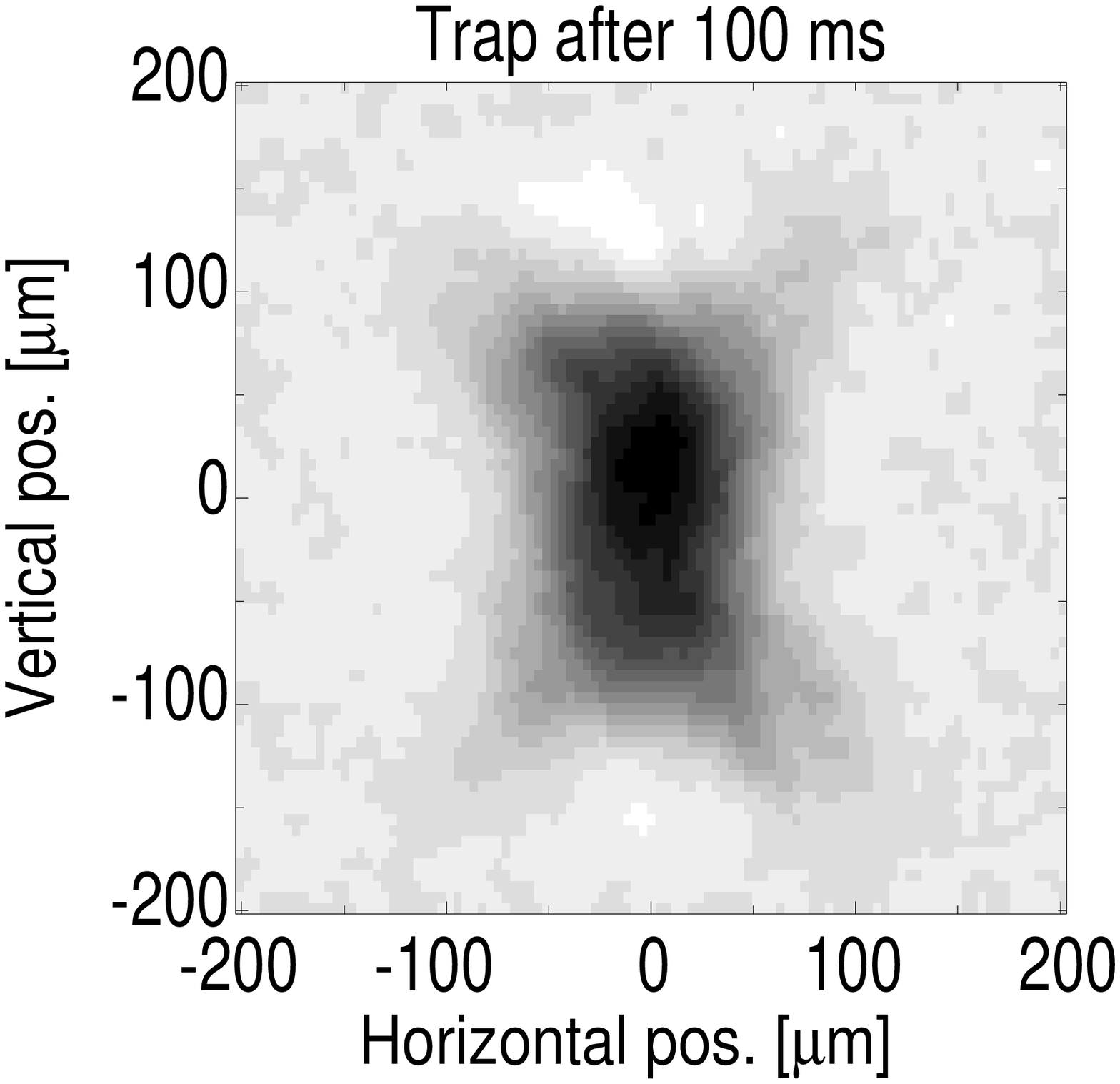}}
\end{picture}
\caption{\footnotesize Absorption images in the lin$||$lin configuration of the YAG beams.
The $X$ shape of the trap which contains $5 \times 10^5$ atoms
is due to an intensity modulation along the vertical (see text).}
\label{linparlin}
\end{figure}

We have observed  that the loading of the YAG trap and the shape of the
atomic cloud was strongly dependent of the polarization of the YAG beams.
When the two YAG beams have linear and parallel polarizations, the number
of trapped atoms is about 5 times higher than in the case of orthogonal
polarizations. The shape of the cloud (fig.\ref{linparlin}) is also very
different as it displays a `X'~pattern.
We attribute this effect as being due to the intensity interference pattern
along the vertical when the two polarizations are parallel. This produces
horizontal planes of micro-wells on  the scale of the YAG wavelength
($\lambda_{\mbox{\scriptsize{YAG}}}/2\sin{53^\circ}$). The atoms are
trapped in these planes of intensity maxima and the effective trapping
volume is larger. An interesting application of this interference effect
might be to enhance the loading efficiency of the YAG trap. After loading
with parallel polarizations we could  rotate the polarization of one of the
beams during the cooling phase to collect these atoms in the bottom of the
harmonic trap.

When the YAG beams have linear and orthogonal polarizations, there is no
intensity modulation but there is a polarization lattice along  $z$ which
leads to a modulation of the light shift potential. It is easy to show that
both of the cesium D1 and D2 transitions (at 894~nm and 852~nm
respectively) contribute to the light shift in a way which depends on the
local polarization. The amplitude of the modulation is proportional to the
frequency difference between the D1 and D2 lines and is opposite for the
$F=3$ and $F=4$ hyperfine states. It represents about $\pm 10\%$ of the
total depth of the potential. Because our YAG laser has several
longitudinal modes, this polarization lattice may fluctuate in time and
produce an unwanted heating.

\subsection{Realization of the Raman excitation profile}

To realize the optimal excitation profile, we first note that the effective
detuning of the Raman transition depends on the power of the Raman beams as
given by eq.(\ref{lightshift}).
\begin{figure}
\centerline{\includegraphics[width= 70mm]{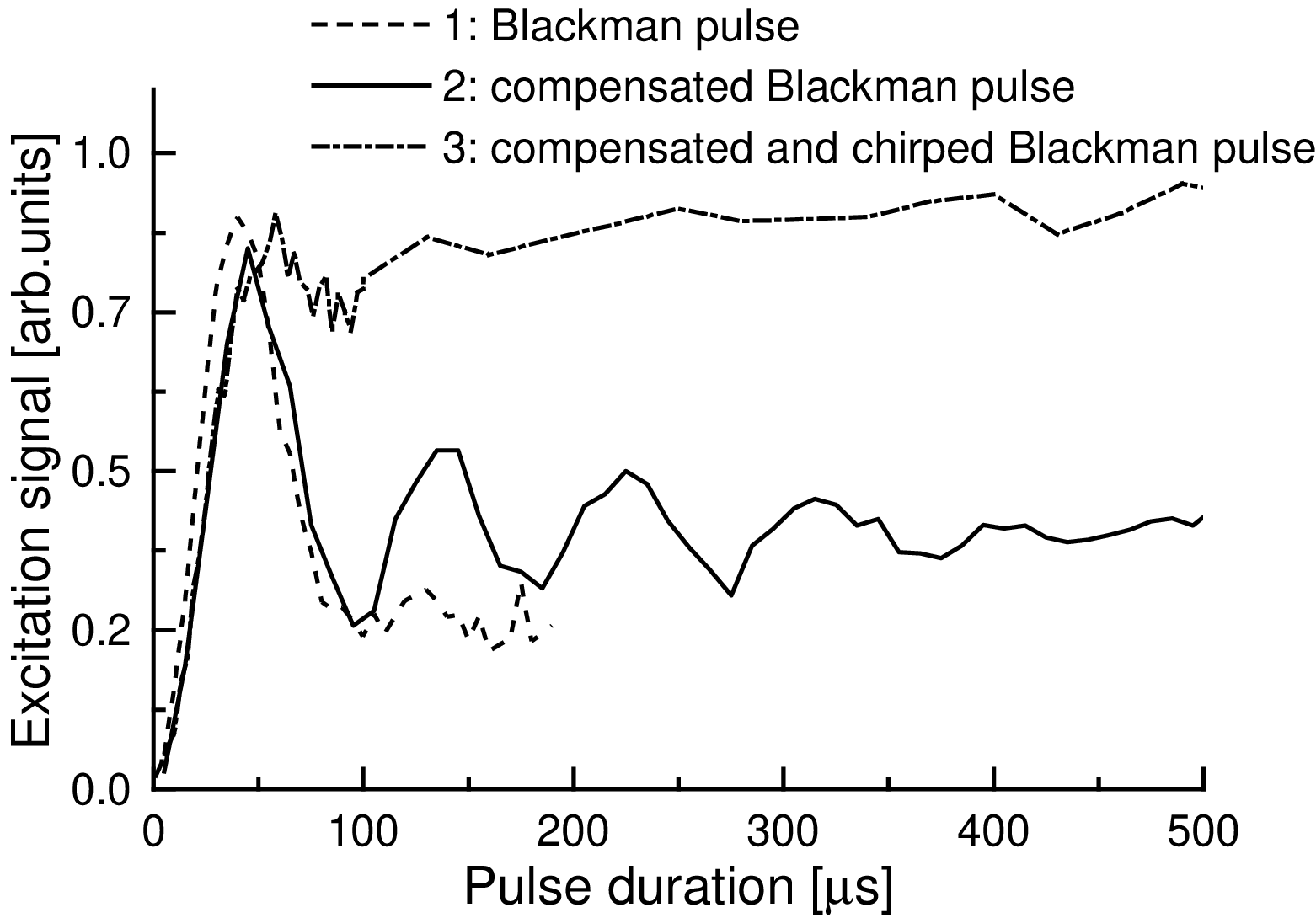}}
\vspace{3mm}
\caption{\footnotesize Excitation rate as a function of the pulse duration; intensity is kept
constant; (1) Blackman pulse without light shift compensation; (2) Blackman
pulse with light shift compensation; (3) chirped Blackman pulse for rapid adiabatic passage
(including light shift compensation).}
\label{Rabi}
\end{figure}

For $\Delta > \delta_{\text{HFS}}$ it is not possible to cancel the light
shift term in eq.(\ref{lightshift}), so that the instantaneous resonance
frequency is shifted proportionally to the instantaneous intensity of the
pulse. This leads to an imperfect $\pi$-pulse condition and to a damping of
Rabi oscillations when using Blackman pulses as shown in fig.\ref{Rabi},
curve 1.

In order to solve this problem, we actively compensate the light shift by
chirping the detuning during the Blackman pulse.  Several periods of the
Rabi oscillation in time become then visible (fig.\ref{Rabi}, curve 2). The
effective Rabi frequency can be measured by this method and for instance
fig.\ref{Rabi} gives $\Omega_{\mbox{\tiny eff}}=7\times 10^4$~rad/s. The
`damping' of the  oscillation is mostly due to the interference between the
various $m$-dependent Rabi frequencies (See eq.(2a)). In the case of the
pulses used for the rapid adiabatic passage which are frequency chirped, we
add the compensation of the light shift to the RAP chirp. In such
conditions and with a frequency chirp symetric around resonance
(fig.\ref{Rabi}, curve 3), the excitation probability remains close to one
for a pulse area greater than $3\pi$.

\begin{figure}
\centerline{\includegraphics[width= 70mm]{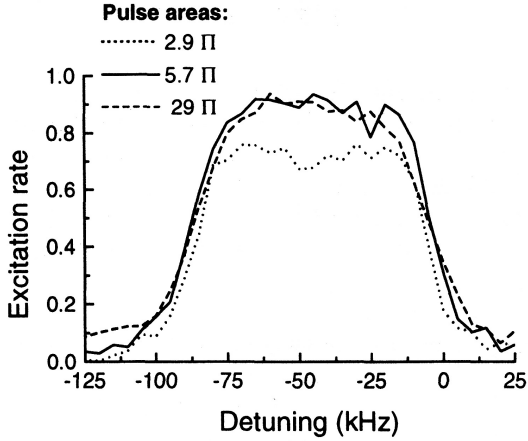}}
\vspace{3mm}
\caption{\footnotesize Excitation probability as a function of detuning for 200~$\mu$s
Blackman pulses chirped from $-81$~kHz to 0; the three curves correspond
to three different pulse areas.}
\label{profile}
\end{figure}

We  then tested the efficiency of the rapid adiabatic passage with
copropagating Raman beams for which there is no velocity selectivity. The
transfer efficiency as a function of Raman detuning $\delta$ is presented
in fig.\ref{profile} for a 200~$\mu$s long Blackman pulse chirped from
$-81$ kHz to 0, $\Delta =2\pi\times 25$~GHz and for various intensities.
The agreement with theory (fig.\ref{transfer}) is excellent.  Note the high
transfer efficiency ($90\%$) and the square shape of the excitation profile
for the pulse of area $5.7 \Pi$ which we will use in the Doppler sensitive
case for Raman cooling (next section).

\subsection{Raman cooling sequence}

For Raman cooling, we use the Raman beams in the counter-propagating
geometry. The transition probability of fig.\ref{profile} becomes a Doppler
sensitive excitation profile: for an atom with velocity $v$ along the Raman
beams, $\delta_{\text{eff}} = \delta + 2 k v$.
\begin{figure}
\centerline{\includegraphics[width= 70mm]{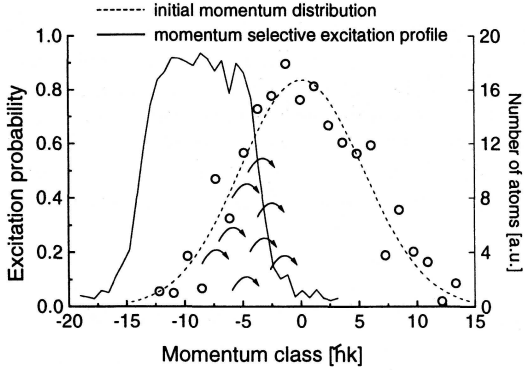}}
\vspace{3mm}
\caption{\footnotesize Measured initial momentum distribution in the crossed dipole trap
($T = 5.6~\mu$K) and Doppler selective excitation profile for a 200~$\mu$s
long chirped Blackman pulse.}
\label{sequence}
\end{figure}
We carefully adjust the initial detuning and the pulse spectral width in
order to avoid any excitation at zero velocity. Fig.\ref{sequence} shows
the momentum dependent excitation profile of a 200~$\mu$s Blackman pulse
chirped over 81~kHz together with the initial momentum distribution. We
optimize the cooling sequence to get the narrowest momentum distribution
after a fixed time (70~ms). We find that the temperature of the cooled
atoms is larger if the pulse rate is too high. This observation is in
agreement with the simple simulation described in section \ref{Xsequence}.
The optimal cooling sequence was: a 300~$\mu$s Blackman pulse exciting the
atoms with momentum between $-12 \hbar k$ and $-3 \hbar k$, followed by a
100~$\mu$s repumping pulse. The total time between two cooling pulses was
then 400~$\mu$s, which corresponds to 7 to 10 pulses per oscillation
period. This sequence was repeated 170 times.

\subsection{Results}

\begin{figure}
\centerline{\includegraphics[width= 70mm]{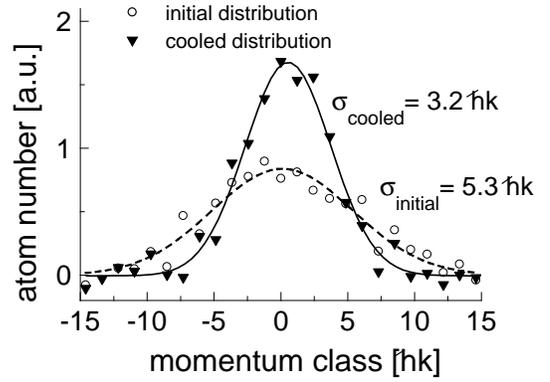}}
\vspace{3mm}
\caption{\footnotesize Momentum distribution before and after Raman cooling.}
\label{cooling}
\end{figure}

The result of the Raman cooling is shown in fig.\ref{cooling}: before
cooling, the momentum distribution along the Raman axis is gaussian with a
$\sigma=5.3 \hbar k$ dispersion. We were able to cool the atomic sample
down to $\sigma=3.2 \hbar k$. This corresponds to a temperature of
$T=2.0~\mu$K. Since the areas of the two momentum distributions are equal,
there is no atom loss during the cooling process.

The momentum dispersion along one axis was reduced by a factor $\eta=1.65$.
The increase in phase space density, proportional to $\eta^6$ for atoms
trapped in an harmonic potential, is a factor 20. We also checked that the
cooling was effective in 3 dimensions by taking absorption pictures of the
trap before and after cooling with the CCD camera. The density of the
cooled atoms is $1.3 \times 10^{12}$~cm$^{-3}$ with an uncertainty of a
factor~2.

\subsection{Limitations and prospects }

For longer cooling sequences, we did not reach a temperature smaller than
2~$\mu$K. Nor was the temperature lower when we added a second, longer,
pulse in order to excite atoms with momentum between $- 3 \hbar k$ and $- 1
\hbar k$. We suspect that Raman cooling is in competition with a residual
heating mechanism due to  a time dependent interference betwen the YAG
beams inducing fluctuations of the trapping potential. A possible solution
to this problem is to alternate in time the two YAG trapping beams at a
rate much higher than the oscillation frequencies. Our numerical
simulations indicate that, after suppression of this heating, temperatures
near the single photon recoil temperature should be accessible. Such a
value would be near the Bose-Einstein condensation threshold, if we neglect
possible density limitations due to photon multiple scattering
\cite{Olshanii95,Cirac96}. A second interesting development is to initiate
evaporative cooling from the high density already obtained in this dipole
trap \cite{Adams95}.

\acknowledgments
We gratefully acknowledge C.Cohen-Tannoudji, J.Dalibard,
J.Reichel, E.Peik, M.Ben  Dahan and Y.Castin for stimulating discussions.
A.\,K. is indebted to the Alexander von Humboldt-Stiftung for support. This
work was supported in part by CNES, NEDO and Coll\`ege de France.
Laboratoire Kastler Brossel is Unit\'e de recherche de l'Ecole Normale
Sup\'erieure et de l'Universit\'e Pierre et Marie Curie, associ\'ee au
CNRS.

\end{document}